\documentclass[amsmath,amssymb,11pt,superscriptaddress,reprint, preprintnumbers, notitlepage,aps,prl,twocolumn, dvipsnames,table]{revtex4-1}
\pdfoutput=1 
\usepackage[utf8]{inputenc}
\usepackage[english]{babel}
\usepackage{amsmath}
\usepackage{graphicx}
\usepackage{dcolumn}
\usepackage{pbox}
\usepackage{amssymb}
\usepackage{epsfig}
\usepackage[dvipsnames]{xcolor}
\usepackage[normalem]{ulem}
\usepackage{slashed}
\usepackage{cancel}
\usepackage{amssymb}
\usepackage{amsmath}
\usepackage{verbatim} 
\usepackage{mathrsfs}
\usepackage{color}
\usepackage{url}
\definecolor{MyDarkBlue}{rgb}{0.1, 0.1, 0.8}
\definecolor{SBlue}{rgb}{0.2, 0.4, 0.7} 
\definecolor{MyLightBlue}{rgb}{0.22,0.51,0.9}
\definecolor{MyGreen}{rgb}{0.0, 0.5, 0.0}
\definecolor{BrickRed}{rgb}{0.8, 0.25, 0.33}
\RequirePackage{hyperref}
\hypersetup{colorlinks, citecolor=SBlue,linkcolor=MyDarkBlue, urlcolor=PineGreen}

\usepackage{comment}

\usepackage{physics}
\usepackage{array}
\usepackage{multirow}
\usepackage{booktabs}
\usepackage{xcolor}

\makeatletter

\makeatletter
\renewcommand\@makecaption[2]{%
  \par
  \vskip\abovecaptionskip
  \begingroup
  
   \small\rmfamily
    \begingroup
     \samepage
     \flushing
     \let\footnote\@footnotemark@gobble
     \@make@capt@title{#1}{#2}\par
    \endgroup
  \endgroup
  \vskip\belowcaptionskip
}
\makeatother

\DeclareUnicodeCharacter{2212}{-}
\begin{document}

\preprint{HRI-RECAPP-2026-05}

\title {Gravitational Wave Imprints of a High-Quality Axion and \\ the Origin of Flavor Hierarchies
}

\author{\bf K.S. Babu}
\email[E-mail:]{babu@okstate.edu}
\affiliation{ Oklahoma State University, Department of Physics, Stillwater, OK 74078, USA}
\author{\bf Sai Charan Chandrasekar}
\email[E-mail:]{sai.sekar@okstate.edu}
\affiliation{ Oklahoma State University, Department of Physics, Stillwater, OK 74078, USA}
\author{\bf Sudip Jana}
\email[E-mail:]{sudip.jana@okstate.edu}
\affiliation{Harish-Chandra Research Institute, Chhatnag Road, Jhunsi, Prayagraj 211019, India}
\affiliation{Homi Bhabha National Institute, Training School Complex, Anushakti Nagar, Mumbai 400094, India}
\author{\bf Sudip Manna}
\email[E-mail:]{sudipmanna@hri.res.in}
\affiliation{Harish-Chandra Research Institute, Chhatnag Road, Jhunsi, Prayagraj 211019, India}
\affiliation{Homi Bhabha National Institute, Training School Complex, Anushakti Nagar, Mumbai 400094, India}
\begin{abstract} Axions, arising from an anomalous global Peccei-Quinn symmetry $U(1)_{\rm PQ}$, offer a compelling solution to the strong CP problem but are vulnerable to Planck-suppressed operators. Gauged abelian flavor symmetries $U(1)_F$, invoked to explain the flavor hierarchies via the Froggatt-Nielsen mechanism, can naturally shield the axion from such effects, yielding an accidental high-quality flavored axion with unit domain wall number. Such constructions predict two complementary signatures: (i) flavor-changing neutral currents from $K\to\pi a$ decays, typically associated with high flavor scales $\Lambda_{\rm FN}\gtrsim f_a$, and (ii) stochastic Gravitational Waves (GWs) sourced by the evolution and decay of gauged flavonic and axionic cosmic-string networks. In addition, global axionic strings can efficiently radiate axions, potentially accounting for the observed dark matter relic abundance. We show that the resulting characteristic plateau--valley structure in the GW spectrum provides a distinctive and powerful probe of high-quality flavored axion dark matter models, complementary to low-energy flavor experiments. 
\end{abstract}

\maketitle

\textbf{\emph{Introduction}.--} Several pressing questions remain unresolved within the Standard Model. One of the most striking is the strong CP problem: the puzzling smallness of the CP-violating parameter $\overline{\theta}$ in QCD stands in sharp contrast to the sizable CP violation observed in the weak sector. Another major drawback is the flavor puzzle, the multiple orders of magnitude spanned by the quark and lepton masses, together with the hierarchical structure of CKM and PMNS mixings, pointing to the need for an underlying theory of flavor. Notably, the solutions to the Strong CP and the flavor problems have been suggested to be closely intertwined~\cite{Davidson:1981zd, Wilczek:1982rv, Babu:1992cu, Ema:2016ops, Calibbi:2016hwq, Babu:2026yqp}.
 
The axion solution to the strong CP problem involves promoting the $\overline{\theta}$ parameter to a dynamical degree of freedom via the Peccei-Quinn (PQ) mechanism~\cite{Peccei:1977hh}. Here, a pseudoscalar ``axion" field arises as the pseudo-Nambu Goldstone Boson (pNGB) of an anomalous global $U(1)_{\text{PQ}}$ symmetry acting on the quark fields~\cite{Weinberg:1977ma, Wilczek:1977pj}. The CP violation arising from non-perturbative QCD instanton effects can be parametrized by the Lagrangian
\begin{equation}\label{eq:axiongluoncoupling}
{\cal L}_{\cancel{\text{CP}}}^{\text{QCD}} = \frac{g_s^2}{32 \pi^2} \left(\overline{\theta}+\frac{a}{f_a}\right) G_{\mu\nu}^a \widetilde{G}^{\mu\nu,a}
\end{equation}
with $\widetilde{G}^{\mu\nu,a} \equiv \frac{1}{2}\epsilon^{\mu\nu\alpha\beta} G_{\alpha \beta}^a$. A potential for the axion is induced via the QCD instantons and can be computed via chiral perturbation theory to be~\cite{Weinberg:1977ma, DiVecchia:1980yfw}
\begin{equation}\label{eq:chiralaxionpotential}
    V(a)\simeq -m_{\pi}^2f_{\pi}^2\sqrt{1-\frac{4m_um_d}{(m_u+m_d)^2}\sin^2{\left(\frac{a}{2f_a}+\frac{\overline{\theta}}{2}\right)}}.
\end{equation}
This potential enjoys a shift symmetry $a\rightarrow a+f_a \overline{\theta}$ with a modified minimum at $\overline{\theta}+a/f_a=0$ where the axion dynamically relaxes the $\overline{\theta}$ parameter to zero, solving the strong CP problem. This, in theory, explains the smallness of the $\overline{\theta}$ parameter in a dynamical fashion where
stringent constraints on $\overline{\theta}\ll10^{-10}$ exist from the non-observation of the electric dipole moment (EDM) of the neutron ~\cite{Crewther:1979pi, Abel:2020pzs}. 

Quantum gravity is expected to violate all global symmetries. Such a solution arising from an anomalous global symmetry is thus susceptible to be broken by quantum gravity effects in the form of Planck-suppressed operators. These operators in turn have couplings that need to be severely fine-tuned (for instance, a $d=5$  operator contributing $\delta f_a^5/M_{\text{Pl}}$ to $\overline{\theta}$ requires its coupling to be $\delta\leq10^{-50}$) in order to be within the observed neutron EDM limits. This is the so called \emph{``axion quality problem"}~\cite{Kamionkowski:1992mf,Holman:1992us,Barr:1992qq}.

Solutions to the axion quality problem often utilize additional, flavor-universal abelian gauge symmetries to stabilize the axion potential against quantum gravitational corrections~\cite{Barr:1992qq, Babu:2024qzb}. A more natural abelian gauge symmetry to consider is a $U(1)_F$ gauged flavor symmetry associated with generating the fermion mass and mixing hierarchies of the SM~\cite{Babu:1992cu, Babu:2026yqp}. 
The flavor hierarchies of the SM fermions can be effectively understood via the Froggatt-Nielsen (FN) mechanism~\cite{Froggatt:1978nt}. Here, all the fermion mass terms and mixings arise as powers of a small parameter $\epsilon$ (typically identified with the Cabibbo angle $\epsilon\sim0.22$), thereby explaining the flavor hierarchy. For this purpose, an SM singlet scalar $X$, the \emph{``flavon"} acquires a Vacuum Expectation Value (VEV) $\langle X\rangle$ that spontaneously breaks the $U(1)_F$ gauge symmetry. 

\begin{figure*}[ht!]
    \centering
    \includegraphics[width=0.7\textwidth]{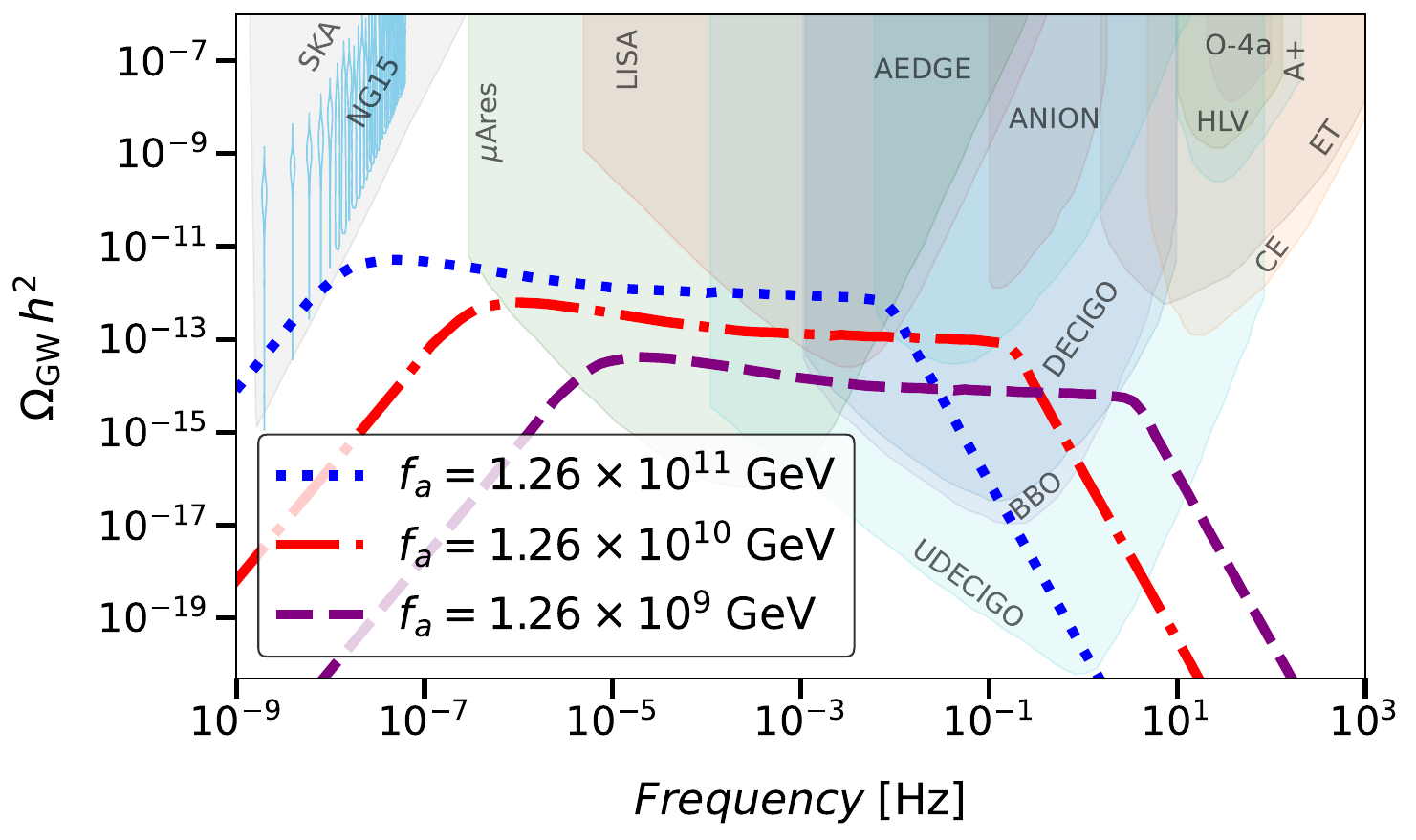}
    \caption{The characteristic plateau-valley GW spectra for different values of axion decay constant $f_a$ at fixed $r=0.6$. The first two benchmark choices, $f_a = 1.26\times10^{11}\,\mathrm{GeV}$ and $f_a = 1.26\times10^{10}\,\mathrm{GeV}$, correspond to the upper and lower edges of the observed axion relic abundance~\cite{Benabou:2024msj, Saikawa:2024bta}, respectively. See text for details. }
    \label{fig:gw_b_plot}
\end{figure*}

\begin{figure*}[htb!]
    \centering
    \includegraphics[width=0.81\textwidth]{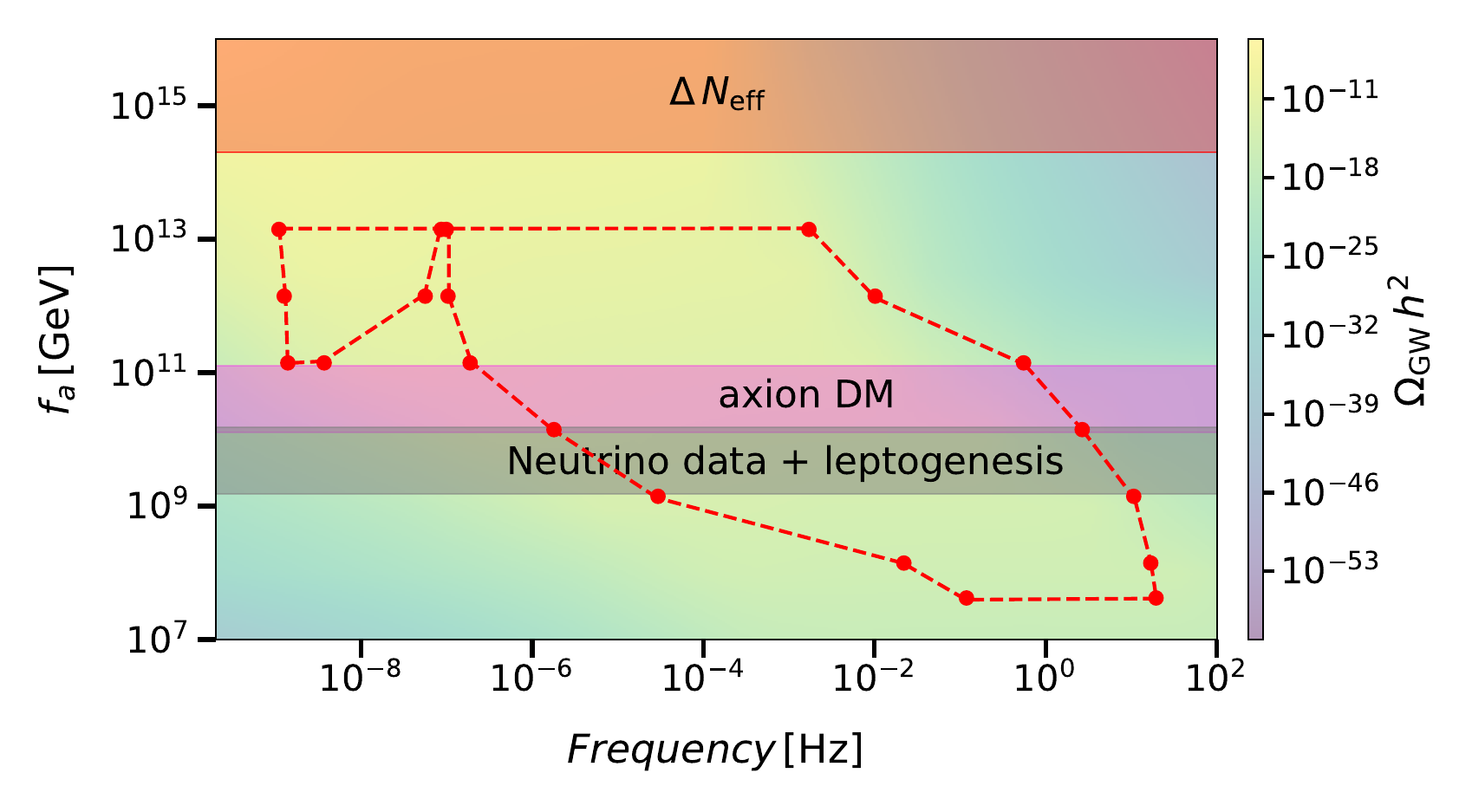}
    \caption{Detection prospects for the corresponding axion decay constant ($f_a$) values across different GW frequency ranges. The combined future sensitivities of the proposed GW experiments-- LISA~\cite{LISA:2017pwj}, $\mu$Ares~\cite{Sesana:2019vho}, BBO~\cite{Harry:2006fi}, DECIGO~\cite{Kawamura:2020pcg}, ET~\cite{Hild:2008ng}, AEDGE~\cite{AEDGE:2019nxb}, and CE~\cite{LIGOScientific:2016wof} are shown by the red dashed contours for a fixed value of $r=0.6$, with the enclosed region being accessible to future GW searches. The magenta band denotes the axion decay constant $f_a$ consistent with the observed DM relic abundance~\cite{Benabou:2024msj, Saikawa:2024bta}, the gray band corresponds to the region favored by the combined fit to neutrino oscillation data and successful leptogenesis, and the red band represents the constraint from the $\Delta N_{\rm eff}$ bound~\cite{Henrot-Versille:2014jua}.
    }
    \label{fig:red_GW_plot}
\end{figure*}

Theories Beyond the Standard Model (BSM) involving additional abelian gauge (global) symmetries spontaneously broken after inflation, commonly predict gauge (global) cosmic strings, one-dimensional cylindrically symmetric topological defects which form networks of long open strings and closed string loops~\cite{Kibble:1976sj, Vilenkin:2000jqa}. These string loops can oscillate and radiate away energy efficiently in the form of GWs from gauge strings~\cite{Vilenkin:1981bx} and Nambu-Goldstone Bosons (NGBs)~\cite{Vachaspati:1984yi}, in this case, the axions, for global strings.  The model of consideration is composed of both global and gauge $U(1)$ symmetries, and there exists a scalar $S$ which is charged under both; therefore, although there exist two kinds of string solutions corresponding to $X=0$ and $S=0$~\cite{Barr:1992qq, Ghosh:2025cxp}, they can emit both GWs and axions as the primary channel of their energy loss mechanism. This complex string structure can thereby lead to some interesting and unique features in the GW spectra, which are shown in  Figs.~\ref{fig:gw_b_plot}-\ref{fig:red_GW_plot} and can be probed in near future by the proposed GW experiments like LISA~\cite{LISA:2017pwj}, $\mu$Ares~\cite{Sesana:2019vho}, BBO~\cite{Harry:2006fi},  DECIGO~\cite{Kawamura:2020pcg}, UDECIGO~\cite{Kudoh:2005as}, Einstein Telescope (ET)~\cite{Hild:2008ng}, AEDGE~\cite{AEDGE:2019nxb}, Cosmic Explorer (CE)~\cite{LIGOScientific:2016wof} and others~\cite{NANOGrav:2023gor, EPTA:2011kjn, EPTA:2023fyk}.
Thus, we find that the characteristic plateau--valley structure in the GW spectrum offers a powerful and distinctive probe of a unified high-quality flavored axion framework capable of simultaneously explaining neutrino masses, the matter--antimatter asymmetry via leptogenesis, fermion flavor hierarchies, the strong CP/axion quality problems, and the observed DM relic abundance~\cite{Babu:2026yqp}.

\vspace{0.1 in}

\textbf{\emph{Flavored accidental axion models}.--}
It was recently worked out in~\cite{Babu:2026yqp}, that the axion can be protected by such a gauged flavor symmetry while naturally explaining the fermion mass hierarchy, resulting in a flavored accidental axion of \emph{``high quality"}, which we term as the \emph{``flaccion"}. The construction requires to generalize the popular DFSZ axion model~\cite{Zhitnitsky:1980tq,Dine:1981rt} utilizing two Higgs doublets $H_u(1,2,\frac{1}{2})$ and $H_d(1,2,-\frac{1}{2})$ and an SM scalar singlet $S$ by the addition of the flavon field $X$. The role of the $S$ field in the DFSZ model is to avoid the Weinberg-Wilczek weak scale axion~\cite{Weinberg:1977ma, Wilczek:1977pj} ruled out by experiments.  This generalization integrates seamlessly with the FN mechanism using two Higgs doublets for the up-type, down-type quarks, plus charged lepton couplings, respectively. By integrating out heavy FN fields, the SM effective Yukawa couplings $y_{ij}^f$ are generated with appropriate suppression of $\epsilon=\langle X\rangle/\Lambda_{\text{FN}}$ based on the $U(1)_F$ charges of the SM fermions $n_{ij}^{u,d,e,\nu,N}$. 
\begin{eqnarray}
{\cal L}_{\rm Yuk} &=& y_{ij}^u Q_i u^c_j H_u \left(\frac{X^{(*)}}{\Lambda_{\rm FN}}\right)^{n^u_{ij}} +  y_{ij}^d Q_i d^c_j H_d \left(\frac{X^{(*)}}{\Lambda_{\rm FN}}\right)^{n^d_{ij}}\nonumber\\ &+&  y_{ij}^\ell L_i e^c_j H_d \left(\frac{X^{(*)}}{\Lambda_{\rm FN}}\right)^{n^\ell_{ij}}+  y_{ij}^{\nu} L_i N_j \widetilde{H_d} \left(\frac{X^{(*)}}{\Lambda_{\rm FN}}\right)^{n^{\nu}_{ij}}\nonumber \\ &+&   M_N N_i N_j \left(\frac{X^{(*)}}{\Lambda_{\rm FN}}\right)^{n^N_{ij}} + \text{H.c}.
\label{eq:Yuk}
\end{eqnarray}
Here, $\Lambda_{\text{FN}}$ is the flavor cutoff scale. For three families ($i=1,2,3$) of SM fermions $Q_i,u_i^c,d_i^c,L_i,e_i^c$ and singlet right-handed neutrinos (RHNs) with flavor charges $q_i,u_i,d_i,l_i,e_i$ and $n_i$ respectively, the Yukawa Lagrangian can be written as

\begin{figure*}[!htb]
    \centering
    \includegraphics[width=0.8\textwidth]{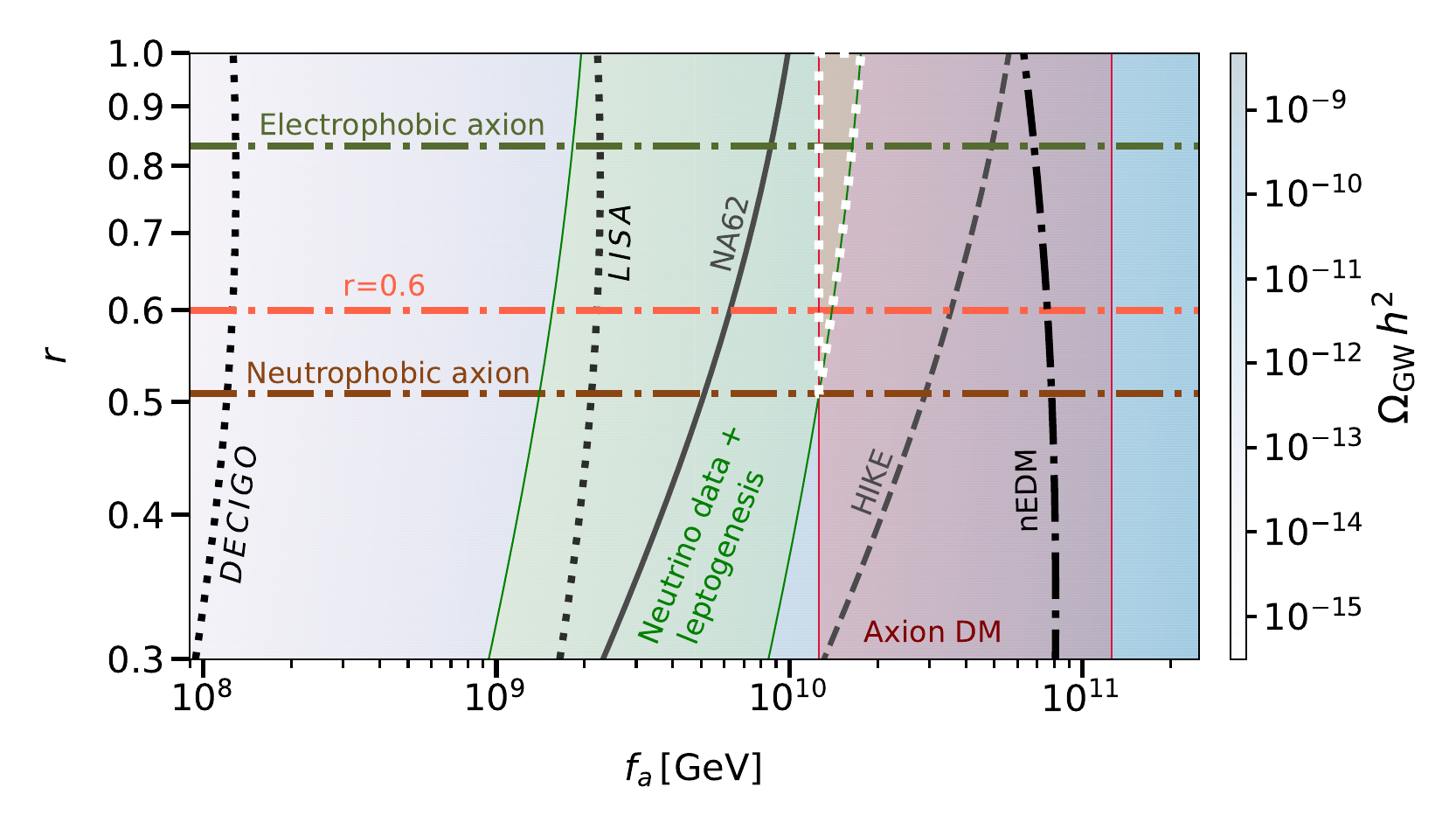}
    \caption{Summary of the model parameter space showing the regions allowed by current experimental and theoretical constraints, together with the projected sensitivities and discovery reaches of future flavor and gravitational-wave observatories.
    The flavor constraints are derived from the current NA62 bounds~\cite{NA62:2025upx} and the projected sensitivity of HIKE~\cite{HIKE:2023ext}, shown by the gray solid and dashed lines, respectively. The neutrophobic and electrophobic axion scenarios are indicated by the brown and olive dot-dashed horizontal lines, respectively. The green band corresponds to the region consistent with neutrino oscillation data and successful leptogenesis, while the light-red band denotes the observed dark matter abundance~\cite{Benabou:2024msj, Saikawa:2024bta}. The requirement of a high-quality axion, inferred from neutron electric dipole moment (nEDM) constraints~\cite{Abel:2020pzs}, excludes the region to the left of the black dot-dashed line. The benchmark choice $r=0.6$ is represented by the orange dot-dashed horizontal line. These constraints are superimposed on the predicted gravitational-wave amplitudes from flavonic and axionic cosmic-string networks, together with the projected sensitivities of the future GW observatories LISA~\cite{LISA:2017pwj} and DECIGO~\cite{Kawamura:2020pcg}, shown as dotted black curves.}
    \label{fig:quality_GW_plot}
\end{figure*}

Here, $\widetilde{H_d} = i \tau_2 H_d^*$ and either $X,X^*$ can couple to each term. In general, $M_N$ is a free parameter associated with the bare mass of the heavy RHNs. Minimally, $M_N$ here is identified with the flavor scale itself: $(M_N)_{ij} = y_{ij}^N\,\Lambda_{\text{FN}}$. An anomaly-free three generation fermion charge assignment that reproduces the full SM fermion mass hierarchy including the neutrinos is $(q_1,q_2,q_3)=(1,0,-2);(u_1,u_2,u_3)=(3,0,-2);(d_1,d_2,d_3)=(-13/3,8/3,8/3)$ for the quarks and $(l_1,l_2,l_3)=(-2/3,16/3,-5/3);(e_1,e_2,e_3)=(-8/3,-8/3,7/3)$. The flavor charges of the scalars $H_u, H_d, X$ are $h_u,h_d,q_X$ respectively and are fixed by the Yukawa interactions in Eq.~(\ref{eq:Yuk}). The RHNs with charges $(n_1,n_2,n_3)=(4,-6,-1)$ naturally cancel the mixed $U(1)_F$ anomalies which also sets up a type-I see-saw scenario~\cite{Minkowski:1977aj,Mohapatra:1979ia} for the generation of light neutrino mass and a ``resonant leptogenesis"~\cite{Pilaftsis:2003gt} scenario in this case~\cite{Babu:2026yqp} for generating the baryon asymmetry of the universe. Both of these features are generally characterized by $M_N\sim\Lambda_{\text{FN}}$, which we hereafter refer to as the ``Neutrino fit and leptogenesis band" in Figs.~\ref{fig:red_GW_plot} and \ref{fig:quality_GW_plot}.

$q_S$, the flavor charge of $S$,  is fixed by the PQ conserving Planck generated interaction term that generates the mass of the pseudoscalar Higgs:
\begin{equation}
V_{\rm{HS}} \supset -\lambda_{\rm{HS}} \frac{H_u H_d S^n (X^{(*)})^k}{M_{\rm Pl}^{n+|k|-2}} + \text{H.c}.
\label{eq:pot1}
\end{equation}
Given an anomaly free flavor charge assignment and normalizing $q_X=1$, values of $(n,k^{\pm})$ decide $q_S$ by gauge invariance, where $\pm k$ denotes whether $X$ or $X^*$ couples to Eq.~(\ref{eq:pot1}).
In this framework, the $U(1)_{\text{PQ}}$ symmetry arises accidentally and the corresponding axion is identified as the orthonormal combination of the phases $\eta_{u},\eta_{d},\eta_{X}$ and $\eta_S$ of the scalars $H_u, H_d, X$ and $S$ respectively as $a=\sum_{\alpha=u,d,X,S} K_\alpha \eta_\alpha$. In the limit $|v_{u,d}| \ll |v_{X,S}|$, the axion decay constant is 
\begin{equation}\label{eq:favX}
f_a \simeq \frac{rq_X v_X 
 }{n\sqrt{q_X^2 +q_S^2r^2}}=\frac{q_X v_S v_X 
 }{n\sqrt{v_X^2q_X^2 +v_S^2q_S^2}},
\end{equation}
where $r=v_S/v_X$ is ratio of the two larger VEVs $\langle S\rangle=v_S$ and $\langle X\rangle=v_X$. The effective scale is defined as $v_{\rm eff}=f_an/q_X$, which will be later used in the GW analysis.

\textbf{\emph{Quality of the axion}.--}
The axion potential in Eq.~(\ref{eq:chiralaxionpotential}) is shifted by contributions from PQ violating operators. The axion that has accidentally arisen is of high quality when this shift is well below the neutron EDM limits, i.e., $\Delta\overline{\theta}<10^{-10}$. The leading PQ violating operators that arise from the loop level (see end matter) are of the form:
\begin{equation}\label{eq:PQbreak}
    V_{\cancel{\text{PQ}}} \supset \Xi~ e^{i\delta }\frac{S^{\mathfrak{a}}X^{\mathfrak{b}}}{M_{\text{Pl}}^{(\mathfrak{a}+\mathfrak{b})-4}}\left( \frac{X}{\Lambda_{\rm FN}}\right)^{\mathfrak{c}}+ \text{H.c}.
\end{equation}
Requirements on the exponents $l,m,\mathfrak{a},\mathfrak{b},\mathfrak{c}$ by gauge invariance and for a high quality axion are $q_S=l/m$, gcd$(l,m)=1$ and $\mathfrak{a}+\mathfrak{b}+\mathfrak{c}=l+m$. Additionally $l,m$ depend on $n,k,h_u,h_d$ from Eq.~(\ref{eq:pot1}). Typically, the first term in Eq.~(\ref{eq:PQbreak}) is safe for axion quality, and the operator deciding the leading shift in $\overline{\theta}$ arises from the loop diagrams involving heavy FN fields for a particular flavor UV completion~\cite{Babu:2026yqp}. This shift can be parametrized in terms of $r,f_a$ and the flavor cutoff $\Lambda_{\text{FN}}$ as
\begin{equation}\label{eq:thetadelta}
\begin{aligned}
\Delta\overline{\theta}_{\rm{loop}}\simeq& \frac{\Xi \sin \delta}{m_{\pi}^2f_{\pi}^2}\frac{(m_u+m_d)^2}{m_um_d} \left| \frac{-3\,n\sqrt{q_X^2 + q_S^2r^2} }{q_X\sqrt{2}}\right|^{\mathfrak{a}+\mathfrak{b}+\mathfrak{c}}\\
&\frac{(f_a)^{\mathfrak{a}+\mathfrak{b}}}{r^{\mathfrak{a}}M_{\text{Pl}}^{\mathfrak{a}+\mathfrak{b}-4}}\left(\frac{f_a}{\Lambda_{\text{FN}}}\right)^{\mathfrak{c}} .
\end{aligned}
\end{equation}
Requiring $\Delta\overline{\theta}_{\rm{loop}}<10^{-10}$ shown as the maroon dot-dashed line in Fig.~\ref{fig:quality_GW_plot}, gives us the viable parameter space $(r,f_a)$ for the high quality axion for a given model characterized by $(n,k)$ in Eq.~(\ref{eq:pot1}) and $q_S$ given in the end matter. 

These models also automatically result in a unit Domain Wall (DW) number
\begin{equation}
 N_{\rm DW}=(q_Xw_1^S-q_Sw_1^X)=1   
\end{equation}
 for 3 generations of SM fermions~\cite{Babu:2026yqp}. Here, $w_1^S,w_1^X\in \mathbb{Z}$ are the winding numbers of the $S$ and $X$ fields respectively that connect to strings attached to a single DW corresponding to a \emph{``unit DW winding axionic string"}. We also note that $q_S, q_X\in \mathbb{Z}$ are the integerized flavor charges of the fields $S$ and $X$ respectively.

\vspace{0.1in}
\textbf{\emph{Flavor phenomenology and astrophysics}.--}
The flavor-dependent couplings of the axion to the SM fermions $\mathcal{L}_{af}=-ia(m_fC_{af}\overline{f}\gamma_5f)/2f_a$ produce flavor violating (FV) neutral meson decays such as $K\rightarrow \pi a$ with the decay width given by
\begin{equation}\label{eq:FV}
\begin{aligned}[b]
\Gamma(K^{\pm} \rightarrow \pi^{\pm}a) &\simeq \frac{m_K^3}{192 \pi f_a^2} \left( 1-\frac{m_{\pi}^2}{m_K^2} \right)^3\\&\abs{\frac{\kappa_{ds}(\epsilon)q_S^2 r^2}{
n\left(q_X^2 + q_S^2 r^2  
 \right)}}^2,
\end{aligned}
\end{equation}
where, $\kappa_{ds}(\epsilon)=\left[(V_d^\dagger P_Q V_d)_{ds} - (V_{d^c}^T P_d V_{d^c}^*)_{ds}\right]\sim\epsilon$ is the off-diagonal mixing parameter of $d,s$ quarks. The NA62~\cite{NA62:2025upx} collaboration which has recently achieved a highly accurate confirmation on $\text{Br}(K\rightarrow\pi \nu \overline{\nu})$ and can project upto $\text{Br}(K^+ \rightarrow \pi^+ a) \lesssim 2.9 \times 10^{-11}$, is the newest candidate for flavored axion searches. Future experiments like HIKE~\cite{HIKE:2023ext} are expected to probe even lower, upto $\text{Br}(K^+ \rightarrow \pi^+ a) \lesssim 1.0 \times 10^{-12}$. Current experiments with similar probes are ORKA~\cite{Worcester:2013aje} and KOTO \cite{KOTO:2024zbl}. Furthermore, for particular values of $r,(n,k),q_S,q_X$, there exist neutrophobic ($C_{an}=0$) and electrophobic ($C_{ae}=0$) points corresponding to various $r$ values, which are highlighted in Fig.~\ref{fig:quality_GW_plot}.

We consider a concrete scenario with $(n,k)=(3,1^+),~h_u=4,~h_d=-2/3,~q_X=1$ fixing $~q_S=-13/9=l/m$ from Eq.~(\ref{eq:pot1}) as a simple and illustrative example and analyze the production of GWs in the coming section, while contrasting it with the axion flavor observables of the current section. Rescaling by $m = 3n$ yields purely integer flavor charges for the relevant fields $q_S=-13,~q_X=9$. For other works on GW probes of FN models from first-order phase transitions and cosmic strings, see refs.~\cite{Ringe:2022rjx, Blasi:2024vew, Antusch:2026msw}.

\textbf{\emph{Gravitational waves from cosmic strings}.--} This class of models predicts two types of cosmic strings corresponding to $S=0$ and $X=0$ string solutions~\cite{Barr:1992qq}, which we refer to as  Type-$a$ and Type-$f$ strings, respectively.

 We further set $r<1$ ($v_S<v_X,v_{\rm eff}\sim v_S$), which allows us to identify Type-$a$ strings as \emph{``axionic"} global strings and Type-$f$ strings as \emph{``flavonic"} gauge strings, and we will follow this convention in our remaining discussions. Therefore, the $X$ field first acquires a VEV at an early time $t_X$ (where the temperature $T\sim v_X$), leading to the formation of Type-$f$ flavonic gauge (local) strings. Subsequently, at a later time $t_S$, the field $S$ acquires a VEV, resulting in the formation of Type-$a$ axionic strings. At the same epoch $t_S$ (where $T\sim v_S$), the pre-existing Type-$f$ flavonic strings adjust their winding numbers to minimize the gradient energy in their vicinity~\cite{Barr:1992qq, Rothstein:1992rh}, which in turn minimizes the number of domain walls expected to be attached to them at $t_{\rm QCD}$, defined as the epoch when the QCD potential for the axion in Eq.~(\ref{eq:chiralaxionpotential}) turns on (when the temperature of the universe is $T_{\rm QCD}\sim\Lambda_{\rm QCD}$, the QCD phase transition temperature). Furthermore, since both fields can carry non-zero windings around the strings and $S$ carries a non-zero charge under the global $U(1)_{\rm PQ}$, Type-$f$ flavonic strings can exhibit an effective global component in addition to their gauge character once $S$ acquires a VEV, leading to an axion emission channel.   

 The fundamental property of a string is its tension $\mu$, which is determined by the winding numbers of the fields around it. We denote the winding numbers associated with the $S$ and $X$ fields around the axionic and flavonic strings by $(w^S_a,w^X_a)$ and $(w^S_f,w^X_f)$, respectively. Here, the $S$ ($X$) superscripts correspond to the winding of the $S$ ($X$) field, and the subscripts $a$($f$) label the axionic (flavonic) strings. 

Therefore, when both $S$ and $X$ acquire VEVs, the tension of the Type-$a(f)$ strings can be written as~\cite{Vilenkin:2000jqa,Cui:2018rwi,Niu:2023khv}
\begin{equation}
\begin{aligned}
  \mu_{a(f)} \simeq\;& \pi \left(w^S_{a(f)}\right)^2 v_S^2 + \pi \left(w^X_{a(f)}\right)^2 v_X^2 \\
  &+ \pi w_{g,a(f)}^2 \, v_{\rm eff}^2 \ln\left(\frac{L(t)}{\delta}\right) \, ,
\end{aligned}
\label{eq.tension_gauge}
\end{equation}
where $w_{g, a(f)} = (q_X w_{a(f)}^S - q_S w_{a(f)}^X)=N_{\rm DW, a(f)}$, denotes the ``axion winding number", corresponding to the long-range Goldstone gradient energy around the Type-$a(f)$ string, which also equals the number of domain walls expected to be attached to a Type-$a(f)$ string once the universe cools to the QCD epoch. $L(t)$ is taken to be of order the Hubble radius, $L(t) \simeq 1/H$, where $H$ denotes the Hubble expansion rate and $\delta\simeq 1/m_{Z'}$ (for $g\sim \mathcal{O}(1)$) corresponds to the string core width, where $m_{Z'}$ is the mass of the heavy gauge boson associated with the gauged flavor symmetry, given by $m_{Z'} = g \sqrt{q_X^2 v_X^2 + q_S^2 v_S^2}$, with gauge coupling $g$. 

Similar to the axion winding number, we can define a ``gauge winding number" $w_{l, a(f)} = (w_1^X w_{a(f)}^S - w_1^S w_{a(f)}^X)$ which corresponds to the number of pure gauge strings attached to no DWs $(N_{\rm DW}=0)$. Such strings have windings labeled, $w_0^S,w_0^X\in \mathbb{Z}$.

For the Type-$a$ strings, initially the configurations with minimal windings $(w^S_{a}=\pm1,\, w^X_{a}=0)$ are expected to be the most abundant~\cite{Barr:1992qq}. Owing to the large logarithmic contribution, the $(w^S_{a}=\pm1,\, w^X_{a}=0)$ strings have global characteristics and predominantly lose energy via axion emission. 
Moreover, we find that these configurations are energetically stable and do not decay into lower axion winding configurations. This can be understood by considering the topologically equivalent decomposition, which follows from charge conservation for cosmic strings, for example~\cite{Mupo:2025ner}:
\begin{equation}
\begin{aligned}
\left(w_{a(f)}^S,w_{a(f)}^X\right)&=|w_{g,a(f)}|(w_1^S,w_1^X)\\
&+|w_{l,a(f)}|(w_0^S,w_0^X).
\end{aligned}\label{eq:decomp}
\end{equation}
So, a $(\pm1,0)$ string decomposes into
\begin{equation}
\label{eq:reorganization_S}
    (\pm1,0)=9(\pm3,\mp2)_1+2(\mp13,\pm9)_0,
\end{equation}
where $0(1)$ corresponds to $N_{\rm DW}=0(1)$ respectively. Although the decomposition of a string into pure gauge strings $(\mp13,\pm9)_0$ with $N_{\rm DW}=0$, and strings $(\pm3,\mp2)_1$ carrying $N_{\rm DW}=1$, is topologically allowed, energetic stability requires that the total string tension be minimized, thereby disfavoring this channel of formation (demonstrated in the end matter).

 In contrast, for the Type-$f$ flavonic strings, configurations with $w^X_{f}=\pm1$ dominate~\cite{Barr:1992qq}. Meanwhile, $w^S_{f}$ adjusts itself to minimize $N_{\rm DW,a(f)}$~\cite{Barr:1992qq,Rothstein:1992rh,Ghosh:2025cxp}, which leads to $w^S_{f}=\mp1$. Similar to the axionic strings, we find that for $\Lambda_{\rm FN} \sim 10^{12}\,\mathrm{GeV}$, upto $r \gtrsim 10^{-7}$, provided the gauge coupling is not extremely small~\cite{Niu:2023khv}, the dominant energy loss channel is still axion emission. This can be understood from the ratio of the GW to axion emission rates, which parametrically scales as~\cite{Chang:2021afa}
\begin{equation}    
\frac{P_{\rm GW}}{P_a} \sim \frac{v_X^4/M_{\rm Pl}^2}{v_S^2}
= \frac{\epsilon^2\Lambda_{\rm FN}^2}{r^2 M_{\rm Pl}^2}\,,
\end{equation}
where the GW and axion emissions are predominantly governed by the gauge and global components of the string tension, respectively. For $\Lambda_{\rm FN} \sim 10^{12}\,\mathrm{GeV}$, this ratio remains subdominant for $r \gtrsim 10^{-7}$ thus demonstrating the axion emission domination. Thus, we observe that the most abundant configuration of flavonic strings corresponds to windings $(w^S_f, w^X_f) = (\pm1, \mp1)$, which are likewise energetically favored, similar to the axionic string configurations discussed earlier. Similar string-wall networks were recently discussed in Ref.~\cite{Bandyopadhyay:2025oju} in the context of high-quality axion dark matter and gravitational waves. The authors considered benchmark values of $r=10^{-2},\,10^{-3}$, and $10^{-4}$ for $f_a=3\times10^{11},\mathrm{GeV}$ and treated the flavonic string as a pure gauge string. In contrast, our analysis indicates that for these benchmark choices, the flavonic string cannot be accurately described by the pure gauge string limit. As a result, the evolution of the string-wall network and the resulting gravitational-wave predictions can differ from those obtained under the pure gauge string approximation.
A topologically equivalent decomposition of these strings into unit axion winding and pure gauge string configurations is given by
\begin{equation}
(\pm1, \mp1) = 4(\mp3, \pm2)_1 + (\pm13, \mp9)_0 \, .
\label{eq:reorganization_X}
\end{equation}
However, this rearrangement is disfavored energetically due to the larger resulting total string tension (see end matter). Therefore, both types of strings follow a similar cosmological evolution up to $t_{\rm QCD}$, with energy loss predominantly governed by axion emission.

Around $t_{\rm QCD}$, both axionic and flavonic strings become attached to multiple domain walls due to their non-trivial axion winding ($N_{\rm DW,a};\,N_{\rm DW,f}>1$). At this stage, the resulting wall-string network is generically stable. As the universe evolves, the energy density of the attached domain walls starts growing and eventually overcomes the energy barrier required for string reorganization. For $m_a$ in the QCD axion range~\cite{Abbott:1982af}, the ratio of the domain wall energy density $\rho_{\rm DW}$ to that of cosmic strings ($\rho_{a(f)}$) at the time of formation (around $t_{\rm QCD}$) can be estimated as
\begin{equation}\label{eq:DWtime}
\frac{\rho_{\rm DW}}{\rho_{\rm str}} \;\sim\; \frac{\sigma H}{\mu_{a(f)}\, H^2}
\;\simeq\; \frac{8 \,m_a\, f_a^2\,M_{\rm Pl}}{\mu_{a(f)} \,\Lambda_{\rm QCD}^2} \, ,
\end{equation}
where $\sigma \simeq 8 m_a f_a^2$ is the surface tension of these domain walls~\cite{Blasi:2023sej}.

For a benchmark point $r = 0.6$, we find that $\rho_{\rm DW}/\rho_{a(f)} > 1$ even at the QCD scale. 
As a result, once domain walls form, both types of strings can fragment into configurations consisting of unit axion winding strings $(\mp3, \pm2)_1$ and pure gauge strings $(\pm13, \mp9)_0$, as shown in Eqs.~(\ref{eq:reorganization_S}) and (\ref{eq:reorganization_X})~\cite{Mupo:2025ner}.

Assuming that this decomposition is the dominant outcome of the network dynamics (since this rearranged configuration is energetically favored at the QCD epoch), the $(\mp3, \pm2)_1$ strings, being topologically unstable, annihilate immediately upon formation primarily via axion radiation, thereby avoiding the domain wall problem.

In contrast, the $(\pm13, \mp9)_0$ pure gauge strings are insensitive to the axion potential and thus decouple from the wall-string network. If these $(\pm13, \mp9)_0$ pure gauge strings do not possess an efficient energy-loss mechanism, their energy density $\rho_{0}$ scales as $1/a^2$ (where $a$ is the scale factor), leading to their domination of the total energy density of the Universe.
However, that scenario is ruled out by Planck~\cite{Planck:2013pxb}, and the only viable alternative is the formation of a horizon-sized string network (composed of string-loops and string-segments) that subsequently decays into GWs~\cite{Kibble:1976sj, Turok:1984db}.

Assuming that about 10\% of the $(\pm13, \mp9)_0$ string loops are large enough to survive against Hubble damping and contribute to GW emission, and presuming that the loop formation rate balances the energy-loss rate of the network, the formation rate of loops of size $\alpha$ is given by~\cite{Vanchurin:2005pa,Olum:2006ix,Martins:2005es,Ringeval:2005kr}
\begin{align}
        \frac{dn_\alpha}{dt}= F_\alpha \frac{\tilde{c}_\alpha}{\alpha}\frac{1}{t^4}.
        \label{eq:string_numer_density}
\end{align}
Based on lattice simulations, the best-fit values of these parameters are: $F_\alpha \simeq \alpha =0.1$, $\tilde{c} \simeq 5.7$ and $0.5$ in the radiation and matter dominated epoch, respectively~\cite{Blanco-Pillado:2017oxo}.

Since these $(\pm13, \mp9)_0$ pure gauged string loops lose energy predominantly through GW emission (primarily via loop oscillations), the instantaneous size of a loop $l(t)$ formed at time $t_i$ with initial size $l_i=\alpha t_i$ can be written as
\begin{align}
        l(t)\simeq\alpha t_i - \Gamma_l G\mu(t-t_i),
        \label{eq:string_ins_length}
\end{align}
where $\mu = \pi \eta^2$~\cite{Vilenkin:2000jqa} is the string tension with $\eta$ the SSB scale ($v_S,v_X$ in this case, since both axionic and flavonic strings can fragment and introduce $(\pm13, \mp9)_0$ gauge strings), and $\Gamma_l \simeq 50$~\cite{Vilenkin:1981bx} is a dimensionless parameter characterizing the GW emission rate.
The total GW radiation from a loop is obtained by summing over all the Fourier frequency modes $\hat{f}_k=2k/\hat{l}(\hat{t})$, where $\hat{l}(\hat{t})$ is the length of the loop at time $\hat{t}$ and $k$ is the mode number.
After accounting for redshift, this frequency is observed today (at $t_0$) as
\begin{align}
    f=\hat{f}_k \frac{a(\hat{t})}{a(t_0)}=\frac{2k}{\hat{l}(\hat{t})}\frac{a(\hat{t})}{a(t_0)}.
\end{align}
Employing Eq.~(\ref{eq:string_numer_density}) and Eq.~(\ref{eq:string_ins_length}) and integrating over the original emission time, the GW-relic today coming from an individual k-th mode of $(\pm13, \mp9)_0$ string oscillation is given by~\cite{Jana:2025vyb}

\begin{align}
\Omega_{\text{GW}}^{(k)}(f) &= 
\frac{1}{\rho_c}\frac{2k}{f}\frac{\Gamma_l^{(k)}G\mu_f^2 F_\alpha}{\alpha\left(\Gamma_l G\mu_f + \alpha\right)}
\int_{t_F}^{t_0} d\hat{t} \, 
\frac{\tilde{c}\left(t_i^{(k)}\right)}{{t_i^{(k)}}^4} \notag \\
&\quad \times
\Theta({t_i^{(k)}}-t_F)
\left(\frac{a(\hat{t})}{a(t_0)}\right)^5
\left(\frac{a(t_i^{(k)})}{a(\hat{t})}\right)^3,
\label{eq:gw_lcs}
\end{align}
where $\rho_c$ denotes the critical energy density. Here, we consider only the contribution coming from flavonic strings, since we are interested in the regime $v_S>v_X,r<1$; therefore, we neglect the contribution coming from axionic strings. The other parameter $t_F$ stands for the time of loop formation, and the other quantity $t_i^{(k)}$ is defined as 
\begin{align}
 t_i^{(k)}(\hat{t},f)=\frac{(\hat{l}(\hat{t},f,k)+\Gamma_l G \mu_f \hat{t})}{( \Gamma_l G \mu_f+\alpha)}.
    \label{Eq:t_hat}
\end{align}
Here, $\Gamma_l^{(k)} \sim k^{-\mathfrak{q}}$, with $\mathfrak{q} = 4/3$ or $2$ corresponding to GW emission from cusps and from kink–kink collisions, respectively.

To evaluate the total GW relic coming from $(\pm13, \mp9)_0$, $\Omega_{\rm GW}^{\rm 0}=\sum_{k=1}^{\infty} \Omega_{\text{GW}}^{(k)}$, one must sum over all $k$ modes in Eq.~(\ref{eq:gw_lcs}). However, the summation becomes numerically expensive for larger values of $k$. We therefore adopt the precomputed spectrum based on Ref.~\cite{Vachaspati:1984gt}, which assumes $\mathfrak{q}=4/3$, and can be found in the \texttt{PTArcade} package~\cite{Mitridate:2023oar}.

In addition, the resulting GW signal can also receive contributions from the decay of the wall-string network. However, since the wall tension in this case scales as $\sigma \sim m_a f_a^2$, the contribution from wall annihilation is subdominant compared to that from cosmic strings. 

The resultant GW spectrum is shown in Fig.~(\ref{fig:gw_b_plot}), where the characteristic transition occurs around the QCD confinement scale $\Lambda_{\rm QCD}$. This frequency scale corresponds to the ultraviolet (UV) cutoff, or equivalently the onset frequency, associated with the QCD epoch around which the $(\pm13,\mp9)_0$ pure gauge strings are formed. The corresponding cutoff frequency can be estimated using the standard frequency-temperature relation~\cite{Cui:2018rwi}, which relates the present-day observed GW frequency to the temperature of the universe at the time of GW emission through cosmological redshift. Since loops produced at different cosmological epochs emit approximately triangular spectra peaked around different characteristic frequencies, the superposition of these redshifted spectra gives rise to the characteristic plateau structure. Consequently, the transition around $f_{\rm cut}$ marks the onset of the plateau-valley behavior in the GW spectrum associated with the formation of the pure gauge string network around the QCD epoch. Using the radiation-dominated relation between cosmic time and temperature, one obtains~\cite{Cui:2018rwi}
\begin{equation}\label{eq:fIRcut}
f_{\rm cut} \simeq \left(8.67 \times 10^{-3} \, \text{Hz}\right)
\left(\frac{T_{\rm QCD}}{\text{GeV}}\right)
\left(\frac{10^{-11}}{G\,\mu_0(v_X)}\right)^{1/2},
\end{equation}
where $G = 1/M_{\rm Pl}^2$ is the gravitational constant and $\mu_0$ denotes the string tension of the $(\pm13,\mp9)_0$ pure gauge strings.

Before the QCD epoch, all strings are predominantly global in nature and primarily lose their energy through axion emission, making the associated GW emission negligible. However, below the frequency scale $f_{\rm cut}$, the GW spectrum begins to rise due to the contribution from loops of the $(\pm13,\mp9)_0$ pure gauge strings formed after the QCD transition.

Therefore, the regime in which the GW amplitude exhibits a sharp dip into the valley around $f_{\rm cut}$ and can serve as a proxy for the QCD scale in relation to the flavor scale $\Lambda_{\rm FN}$, since $v_X=\epsilon \,\Lambda_{\rm FN}$ from Eq.~(\ref{eq:fIRcut}). This feature can provide a novel signature and a potential avenue to illuminate the QCD scale and low energy flavor observables through GW observations. This feature in the mHz frequency range can be probed by LISA~\cite{LISA:2017pwj}, which is expected to begin science operations around 2036,  as well as DECIGO~\cite{Kawamura:2020pcg}, BBO~\cite{Harry:2006fi}, and UDECIGO~\cite{Kudoh:2005as}, which are designed to achieve even higher sensitivities around a similar frequency range and are shown in Fig.~(\ref{fig:red_GW_plot}). Finally, the combined sensitivities of the flavor and GW experiments, along with the preferred region solving the five major SM puzzles (dotted in white), are shown in Fig.~(\ref{fig:quality_GW_plot}). The white dotted region for such flaccion models points to complementary probing by both future flavor experiments like HIKE~\cite{HIKE:2023ext} and future GW detectors like LISA~\cite{LISA:2017pwj} by narrowing down the $\Lambda_{\rm FN}$ value within the preferred band.

We finally comment on the production of axion DM satisfying the observed relic density in the post-inflationary regime. When $t_{\rm DW}\sim t_{\rm QCD}$ where $t_{\rm DW}$ is the collapsing time of the string wall network obtained from Eq.~(\ref{eq:DWtime}), the axion emission is dominated by the decay of global strings for $r\gtrsim10^{-3}$ which favors an axion parameter window of $m_a\in[45,450]\,\,\mu\text{eV}, f_a\in1.27\times[ 10^{10},10^{11}]\,\,\text{GeV}$~\cite{Benabou:2024msj, Saikawa:2024bta} which we have represented as the axion DM bands in Figs.~\ref{fig:gw_b_plot}-\ref{fig:quality_GW_plot}. For $r\lesssim10^{-3}$, the unstable domain wall network contribution needs to be considered for the total axion DM production, which changes the DM band, which appears to increase the window to about $100~\text{meV}$~\cite{Niu:2023khv}, thereby expanding the parameter space of axion DM. All of these are accounted for in the combined plot of Fig.~\ref{fig:quality_GW_plot}, demonstrating future complementary sensitivities of low-energy flavor experiments like HIKE~\cite{HIKE:2023ext} and GW experiments like LISA~\cite{LISA:2017pwj}. 

\vspace{0.1 in}

\textbf{\emph{Conclusions}.--} We have shown that gauged flavor symmetries can simultaneously address fermion mass hierarchies and solve the axion quality problem, leading to a predictive cosmic-string network and an associated stochastic GW signal. The resulting GW spectrum, characterized by a distinctive plateau--valley structure, furnishes a sharp and testable imprint of flavored accidental axion frameworks, directly linking the strong CP and flavor puzzles. Upcoming GW observations thus open a decisive window on axion quality and high-scale flavor structure, contrasting low-energy flavor experiment probes. Strikingly, this framework simultaneously addresses five pressing puzzles of particle physics, neutrino masses, matter-antimatter asymmetry via leptogenesis, fermion flavor hierarchies, the strong CP/axion quality problems, and dark matter, while predicting a distinctive plateau--valley GW spectral signature as its experimental hallmark.

\vspace{0.2 in}
\begin{acknowledgments}

{\textbf {\textit {Acknowledgments.--}}}
The work of KSB and SCC is supported in part by the US Department of Energy grant number DE-SC0016013. SJ and SM would like to acknowledge the support from the Department of Atomic Energy (DAE), Government of India.

\end{acknowledgments}


\section*{End Matter}

\textbf{\emph{Energetically favorable configuration of axionic strings}.--}
For $(w_a^S,w_a^X)=(\pm 1,0), m=q_X=9,\, l=q_S=-13$, the number of  unit axion winding strings and number of pure gauge strings can be obtained from Eq.~(\ref{eq:decomp}) as
\begin{equation}
\begin{aligned}[b]    
\big|w_{g,a}\big|_{(\pm 1,0)}&= 9,\\
\big|w_{l,a}\big|_{(\pm 1,0)}&=w_1^S w_a^X-w_1^Xw_a^S= 2 .
\end{aligned}
\end{equation}
The resultant string tension of this winding configuration from Eq.~(\ref{eq.tension_gauge}) in terms of $f_a$ is 
\begin{equation}
\mu_a\Big|_{(\pm 1,0)} \simeq  \frac{\pi}{9}\, f_a^2+ \frac{\pi}{9}  \, f_a^2 \ln\left(\frac{L(t)}{\delta}\right).
\label{eq.tension_S_app}
\end{equation}
Since $\ln\left(L(t)/\delta\right)\sim\mathcal{O}(10^2)$~\cite{Chang:2021afa}, the IR logarithmic term dominates the axionic string tension. Now, after reorganization the $(\pm 1,0)$ strings breaks up as
\begin{equation}
\begin{aligned}[b]
(\pm 1,0) &= \big|w_{g,a}\big|_{(\pm 1,0)}(\pm 3,\mp 2)_1\\& + \big|w_{l,a}\big|_{(\pm 1,0)}(\mp 13,\pm 9)_0\, .
\end{aligned}
\end{equation}
For $(\pm 3,\mp 2)$, we find $\big|w_{g,a}\big|_{(\pm 3,\mp 2)}=N_{\rm DW} =1$ and for $(\mp 13,\pm 9)$, $\big|w_{g,a}\big|_{(\mp 13,\pm 9)}=N_{\rm DW}=0$ as constructed.
Now, the total string tension of the reorganized state composed of nine  $(\pm 3,\mp 2)_a$ unit axionic and two $(\mp 13,\pm 9)_f$ pure gauged strings in terms of the flavor scale $\Lambda_{\rm FN}$ and $f_a$ is
\begin{equation}
\mu_a\Big|_{\rm Re} \simeq 198\pi\, (\epsilon \Lambda_{\rm FN})^2+419\pi \, \frac{f_a^2}{9}+\pi  \, f_a^2 \ln\left(\frac{L(t)}{\delta}\right).
\end{equation}
It is readily apparent that $\mu_a\Big|_{\rm Re} > \mu_a\Big|_{(\pm 1,0)}$, demonstrating that the most abundant $(\pm 1,0)$ axionic configurations are energetically favored over the reorganized pure $[9(\pm 3,\mp 2)_1 + 2(\mp 13,\pm 9)_0]$ configurations.


\textbf{\emph{Energetically favorable configuration of flavonic strings}.--}
Similarly, for $(w_f^S,w_f^X)=(\pm 1,\mp1)$, the number of unit axion winding strings and pure gauge strings is
\begin{equation}
\big| w_{g,f}\big|_{(\pm 1,\mp1)} =4, \big| w_{l,f}\big|_{(\pm 1,\mp1)}=w_1^S w_f^X-w_1^Xw_f^S =1\, .
\end{equation}
This results in a string tension of
\begin{equation}
\begin{aligned}[b]
\mu_f\Big|_{(\pm 1,\mp1)} &\simeq \pi\,(\epsilon \Lambda_{\rm FN})^2+\pi\,\frac{f_a^2}{9} \\&+ 16\pi \, \, \frac{f_a^2}{9} \ln\left(\frac{L(t)}{\delta}\right),
\label{eq.tension_S_app}
\end{aligned}
\end{equation}
This, in turn, leads to a reorganized combination
\begin{equation}
\begin{aligned}[b]
(\pm1, \mp1) &= \big| w_{g,f}\big|_{(\pm 1,\mp1)}(\mp3, \pm2)_1 \\&+ \big| w_{l,f}\big|_{(\pm 1,\mp1)}(\pm13, \mp9)_0\, .
\end{aligned}
\end{equation}
whose resultant total string tension composed of four pure $(\pm3,\mp2)_1$ unit winding axionic and one $(\mp 13,\pm 9)_0$ pure gauge string is
\begin{equation}
\begin{aligned}[b]
\mu_f\Big|_{\rm Re} &\simeq 97\pi\, (\epsilon \Lambda_{\rm FN})^2+205\pi \,\frac{f_a^2}{9}\\&+ 4\pi  \, \frac{f_a^2}{9} \ln\left(\frac{L(t)}{\delta}\right),    
\end{aligned}
\end{equation}
since $\ln\left(L(t)/\delta\right)\sim\mathcal{O}(10^2)$~\cite{Chang:2021afa}, we have $\mu_f\Big|_{\rm Re} > \mu_f\Big|_{(\pm 1,\mp 1)}$. Therefore, the most abundant $(\pm 1,\mp 1)$ flavonic string configurations are energetically favored over the reorganized pure configurations $4(\pm 3,\mp 2)_1 + (\mp 13,\pm 9)_0$. This energy barrier of the reorganized strings is thus present for all flavored accidental axion models due to the presence of the possible decomposition for this class of high-quality axion models.

\textbf{\emph{nEDM bounds for high quality axion}.--}
For $(n,k)=(3,1^+),~h_u=4,~h_d=-2/3,~q_X=1,~q_S=-13/9$ and the specific UV completion given in Ref.~\cite{Babu:2026yqp} the most dominant operator that contributes to nEDM is
\begin{equation}
V_{\cancel{\text{PQ}}}^{\text{loop}}\supset \left( \frac{\ln (\Lambda_{\rm FN}/M_{\text{A}_\text{H}})}{16\pi^2}\right)^2\frac{S^{9}X^2}{M_{\rm Pl}^7}\left( \frac{X}{\Lambda_{\rm FN}}\right)^{11}+ \text{h.c.}
\label{eq:nEDM}
\end{equation}
This arises from a two-loop diagram involving the heavy FN fields as internal lines and is more leading than the naive operator~\cite{Babu:2026yqp}
\begin{equation}
V_{\cancel{\text{PQ}}}\supset\frac{S^9 X^{13}}{M_{\rm Pl}^{18}}+\text{h.c}
\label{eq:nEDM_naive}
\end{equation}
Using Eq.~(\ref{eq:thetadelta}) with $(\mathfrak{a},\mathfrak{b},\mathfrak{c})=(9,2,11)$ arising from Eq.~(\ref{eq:nEDM}) and requiring $\Delta\theta_{\text{loop}}<10^{-10}$ gives the axion quality bound in Fig.~(\ref{fig:quality_GW_plot}).

\bibliographystyle{utphys}
\bibliography{reference}
\end{document}